\documentclass{ws-procs9x6}
\begin{document}

\title{DIFFUSION MONTE CARLO CALCULATIONS FOR\\
       THE GROUND STATES OF ATOMS AND IONS IN\\
NEUTRON STAR MAGNETIC FIELDS}

\author{S. B\"UCHELER, D. ENGEL, J. MAIN and G. WUNNER$^*$}

\address{Institut f\"ur Theoretische Physik 1, Universit\"at Stuttgart\\
70550 Stuttgart, Germany\\
$^*$E-mail: wunner@itp1.uni-stuttgart.de\\
}

\begin{abstract}
The  diffusion quantum Monte Carlo method is 
extended to solve the old  theoretical physics problem of many-electron
atoms and ions in intense magnetic fields. The feature of our approach is the 
use  of  adiabatic approximation wave functions augmented by a Jastrow factor
as guiding functions to initialize the quantum Monte Carlo prodecure. We calculate the ground state energies of atoms and ions with nuclear charges from $Z= 2, 3, 4, \dots, 26$
for  magnetic field strengths relevant for 
neutron stars. 
\end{abstract}


\bodymatter

\section{Introduction}\label{aba:sec1}

The  discovery of features in the X-ray spectra of the thermal emission spectra of the isolated neutron star {1E 1207} 
\cite{sanwal2002,mereghetti2002} and three other isolated neutron stars 
has revived the interest in studies of medium-$Z$ elements in strong magnetic fields.
The reason is that the  observed features could be due to atomic transitions in elements that are fusion products of the progenitor star.  However, to calculate synthetic spectra for 
model atmospheres, and thus to be in a position to draw reliable conclusions from observed spectra to the elemental composition of the atmosphere and the distribution  of elements on different ionization stages, accurate atomic data for these elements at very strong magnetic fields ($\sim 10^7$ to $10^9$~T) are indispensible.

While the atomic properties of hydrogen and, partly, helium at such field
strengths have been clarified in the literature over the last 25 years
(for a detailed list of references see, e.~g., Ref.~\refcite{buecheler07}), for elements with nuclear charges $Z > 2$ only fragmentary atomic data exist 
with an accuracy necessary for the calculations of synthetic spectra. 

We have tackled \cite{buecheler07} the problem by adapting
the diffusion Monte-Carlo method (DQMC) \cite{reynolds82,bernu88,jones97}
to the case of neutron star magnetic fields. This method has the
advantage that ground-state energies can be determined practically
free from approximations. 
\section{DQMC for neutron star magnetic fields}

The basic idea of DQMC  is to 
identify the ground state  wave function  {$\Phi_0\,(\vec R,t)$} ($\vec R = (\vec r_1, \dots, \vec r_N))$ of an $N$-body Hamiltonian ${\hat H}$ 
with a  {\em particle density} whose correct distribution is found by following the
random walk of many test particles ("walkers") in imaginary time  in  3{$N$}-dimensional configuration space. To reduce fluctuations one works with
a density distribution $f(\vec R,\tau) \equiv  \Psi(\vec R,\tau) \Psi_{\rm G}(\vec R)$, where  $\Psi_{\rm G}$ is a given guiding function used for importance sampling. The density distribution $f$  
obeys a drift-diffusion equation in imaginary time. Because the importance-sampled Green's function is an exponential operator, one can expand it in terms of a Euclidean path integral. For sufficiently small time steps
one can write down accurate approximations to the Green's function, and sample
it with diffusion Monte-Carlo \cite{buecheler07,reynolds82,bernu88,jones97}.

\subsection{Choice of the guiding functions}
The choice of the guiding function is crucial for the success of the DQMC procedure. We take the guiding function $\Psi^{\rm ad}_{\rm G}$ as a Slater determinant of single-particle
orbitals each of which is a product of a Landau state in the lowest level with a given magnetic quantum number and an unkown longitudinal wave function
("adiabatic approximation" \cite{schiff37}). The different longitudinal
wave functions are obtained selfconsistently by an iterative solution of the Hartree-Fock equations using B-splines on finite elements.

\subsection{Jastrow factor}
To incorporate correlation effects it is common to multiply the
guiding function by a Jastrow factor, $\Psi_{\rm G}=\Psi^{\rm JF}\Psi^{\rm ad}_{\rm G}={\rm e}^{-U(\vec R)}\Psi^{\rm ad}_{\rm G}$. We adopt the form
\begin{equation}
U=  - {1}/{4} \, \sum_{i<j}^N{r_{ij}}/(1+\sqrt{\beta}\,r_{ij})+ Z\sum_{i=1}^N{r_i}/(1+\sqrt{\beta}\,r_i)~,
\end{equation}
where $\beta$ is the magnetic field strength in atomic units ($\beta = B/B_0$,
$B_0 = 4.701 \times 10^5$ T). This leads to modifications of the adiabatic approximation guiding functions only 
at small distances, of the order of the Larmor radius.

\section{Results and Discussion}

\begin{figure}
\begin{center}
\includegraphics[width=4.5in]{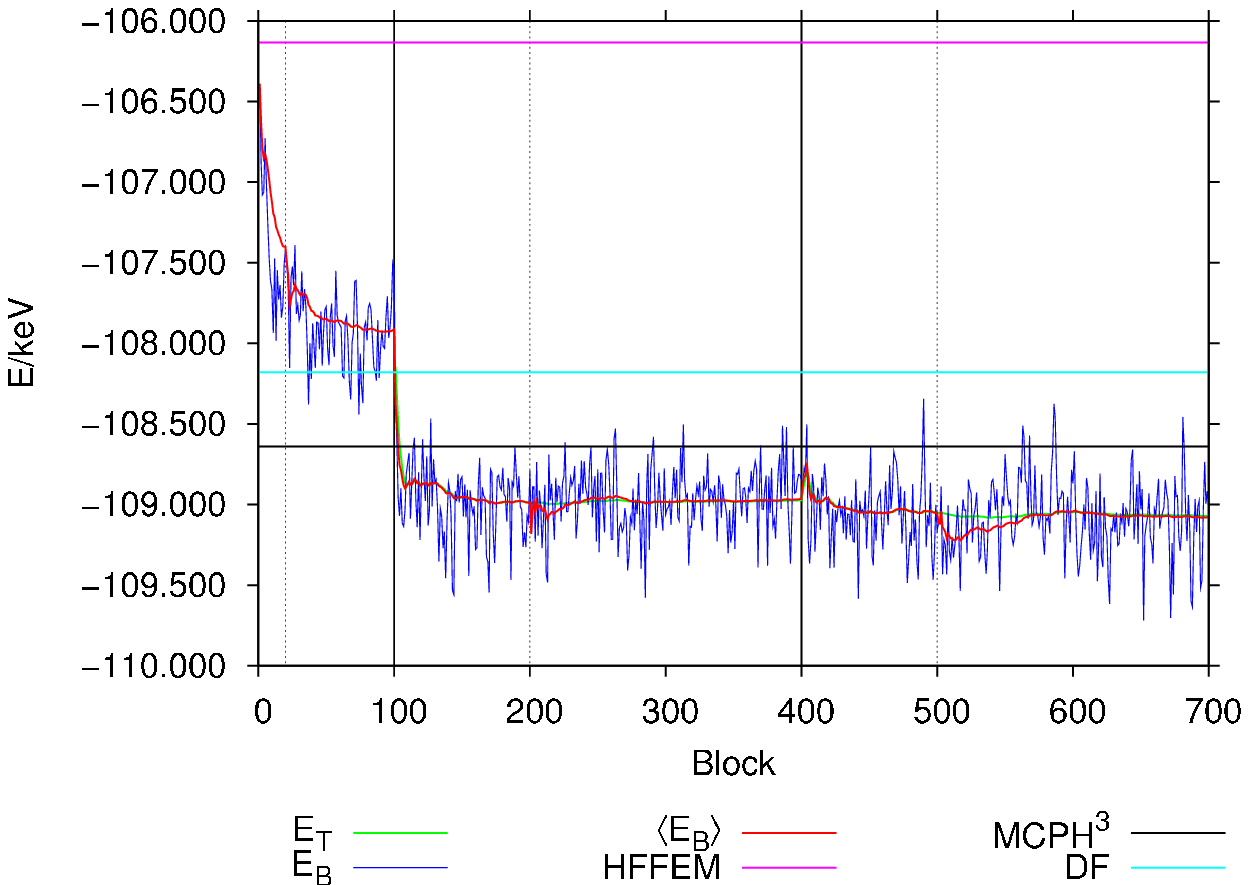}
\end{center}
\caption{ Behavior of the block energy $E_{\rm B}$ (ragged curve) and the 
averaged block energy $\langle E_{\rm B}\rangle$ (smooth curve) in the DQMC simulation for 
the ground state energy 
of neutral iron ($Z = 26$) at $B = 5 \times 10^8$~T as a function of the number of blocks.  In each block, 200 time steps $\Delta\tau= 5 \times 10^{-6}$\,a.u. were performed. 
 (HFFEM (top horizontal line): energy value in adiabatic approximation;
DF (second horizontal line from top): density functional result of 
Ref~\refcite{jones85}; 
MCPH$^3$ (third horizontal line from top): result of Ref.~\refcite{mori2002}.)
}
\label{fig1}
\end{figure}

As a representative example, Fig.~\ref{fig1} shows for the ground state of  neutral iron ($Z = 26$) at $B = 5 \times 10^8$~T the typical flow of a diffusion quantum Monte Carlo simulation. Ions can be treated without additional complication in the same way \cite{buecheler07}. The figure depicts the energy offset $E_ {\rm T}$, the block energy $E_{\rm B}$ and the averaged block energy $\langle E_{\rm B}\rangle$ as a function of the number of blocks performed. 

The complete simulation goes through three stages. During the first 100 blocks, a variational quantum Monte Carlo calculation (VQMC) is performed. Since the adiabatic approximation guiding wave function is augmented by the Jastrow factor, the VQMC calculation already lowers the energy in comparison with the initial adiabatic approximation result. This stage is followed, in the next 300 blocks, by a fixed-phase diffusion quantum Monte Carlo (FPDQMC) simulation. It is seen that the onset of the simulation leads to a considerable drop in the energy. Finally, in the last 300 blocks a released-phase diffusion quantum Monte Carlo (RPDQMC) simulation is carried out, which still slightly lowers the averaged block energy, by roughly
0.1 per cent. The dashed vertical lines in Fig.~\ref{fig1} indicate the blocks where dynamical equilibrium of the walkers
is reached. The relatively small difference between the fixed-phase and the released-phase results indicates that the phase of the adiabatic approximation wave function already well reproduces the phase of the ground state wave function. The small fluctuations of the individual block energies $E_{\rm B}$ evident in Fig.~\ref{fig1} are characteristic of diffusion quantum Monte Carlo simulations. It is also seen, however, that the averaged block energies $\langle E_{\rm B} \rangle $ quickly converge to  constant values in all three stages of the simulation. 
Our final RPDQMC result for the energy is 
$E_0 = -109.079$  keV and lies well below the density functional (DF) value. The standard 
deviation of the block energies at the end of the simulation in this case 
is $\sigma = \pm 0.186$ keV.

\begin{sidewaystable}
\tbl{Energy values in keV for the ground states  from helium to iron   at  $B=10^8$~T. Parameters of the QMC simulations: 500 walkers, time steps  $\Delta\tau (Z=2, \dots, 10)=10^{-4}\mathrm{\,a.u.}$, $\Delta\tau (Z=11, \dots, 19)=5\times10^{-5}\mathrm{\,a.u.}$, $\Delta\tau (Z=20, \dots, 26)=2\times 10^{-5}\mathrm{\,a.u.}$ (discussion see text).}
{\begin{tabular}{cD{.}{.}{6}D{.}{.}{6}D{.}{.}{6}D{.}{.}{6}D{.}{.}{6}D{.}{.}{6}D{.}{.}{6}}\toprule
 Z&\multicolumn{1}{c}{RPDQMC}&\multicolumn{1}{c}{FPDQMC}&\multicolumn{1}{c}{VQMC}&\multicolumn{1}{c}{HFFEM}&\multicolumn{1}{c}{2DHF}&\multicolumn{1}{c}{MCPH$
^3$}&\multicolumn{1}{c}{DF}\\\colrule
2& -0.5827&-0.5827&-0.5791&-0.5754&-0.57999&-0.5766&-0.6035^{\text b}\\
 3& -1.230&-1.229&-1.220&-1.211&-1.22443&-1.214&\\
 4& -2.081&-2.080&-2.065&-2.044&-2.07309&-2.056&\\
 5& -3.122&-3.119&-3.095&-3.057&-3.10924&-3.085&\\
 6& -4.338&-4.331&-4.294&-4.236&-4.31991&-4.288&-4.341^{\text b}\\
 7& -5.716&-5.712&-5.660&-5.568&-5.69465&-5.657&\\
 8& -7.252&-7.246&-7.173&-7.045&-7.22492&-7.176&\\
 9& -8.938&-8.930&-8.834&-8.658&-8.90360&-8.845&\\
10&-10.766&-10.753&-10.630&-10.400&-10.72452&-10.664&-10.70^{\text a}\\
11&-12.725&-12.716&-12.569&-12.266&&-12.625&\\
12&-14.827&-14.817&-14.618&-14.249&&-14.745&\\
13&-17.061&-17.043&-16.813 &-16.352 [1]&&-16.973 [1]&\\
14&-19.480&-19.461&-19.185 &-18.619 [1]&&-19.408 [1]&-19.09^{\text a}\\
15&-22.022&-22.009&-21.665 &-21.002 [1]&&-21.987 [1]&\\
16&-24.700&-24.668&-24.275 &-23.482 [2]&&-24.718 [2]&\\
17&-27.541&-27.523&-27.044 &-26.130 [2]&&-27.618 [2]&\\
18&-30.529&-30.509&-29.950 &-28.890 [2]&&-30.766 [2]&\\
19&-33.650&-33.605&-32.999 &-31.756 [2]&&-34.036 [2]&\\
20&-36.891&-36.881&-36.145 &-34.750 [3]&&-37.500 [3]&-35.48^{\text a}\\
21&-40.296&-40.274&-39.458 &-37.865 [3]&&&\\
22&-43.867&-43.821&-42.900 &-41.083 [3]&&&\\
23&-47.526&-47.490&-46.458 &-44.426 [4]&&&\\
24&-51.360&-51.271&-50.102 &-47.877 [4]&&&\\
25&-55.279&-55.224&-53.915 &-51.430 [5]&&&\\
26&-59.366&-59.311&-57.913 &-55.108 [5]&&&-56.01^{\text a}\\\botrule
\end{tabular}}
\begin{tabnote}
$^{\text a} $Ref.~\refcite{jones85}.
\quad$^{\text b}$Ref.~\refcite{medin2006}.
\end{tabnote}\label{Table3}
\end{sidewaystable}

Table~\ref{Table3}  lists the results for all elements from helium to iron at the magnetic field strength  $B=10^8$~T. 
The table contains in the first three columns the results of the three stages of the simulation and in the fourth column the energy values in adiabatic approximation calculated with our own Hartree-Fock finite-element method
HFFEM. Literature values obtained by Ivanov and Schmelcher \cite{ivanov00} (2DHF), by Mori and Hailey \cite{mori2002} (MCPH$^3$, multi-configurational perturbative hybrid Hartree-Hartree-Fock)  and the results of density functional calculations
\cite{jones85,medin2006} (DF) are given in the remaining columns. The numbers in brackets attached to the HFFEM, 2DHF, 
MCPH$^3$ and DF results designate the number of electrons occupying an excited hydrogen-like single-particle longitudinal state. 
It can be seen 
that already the fixed-phase results lie slightly below the values that were obtained using the 2DHF method. The comparison with the results of the MCPH$^3$ method shows that our RPDQMC energy values generally lie below those results, but there are also exceptions where our results lie above the MCPH$^3$ energies. This may be due to the fact that the hybrid method is not 
self-consistent, since it evaluates the exchange energy in first-order perturbation theory in a basis of Hartree states and it does not include the back-reaction of the excited Landau states whose admixtures are taken into account perturbatively on the effective interaction potentials. Therefore the method need not necessarily produce an upper bound on the energy.

The comparison with the results of the DF calculations shows that these yield lower ground state energies at small nuclear charge numbers than our RPDQMC results, while for large $Z$ the reverse is the case. The DF results listed in Table~\ref{Table3}  differ in the choice of the exchange functional. Given this restriction, it cannot be ensured that the DF calculations in all cases produce an upper bound on the ground state energy in magnetic fields as do the ab-initio methods used in this work or in a Ref.~\refcite{ivanov00}.
\section{Conclusions}

We have extendend the 
released-phase diffusion Monte Carlo method to the calculation of the
ground state energies of  atoms and ions from helium to iron neutron star magnetic field strengths by using adiabatic approximation
wave functions as guiding wave functions \cite{buecheler07}. However, for matching observed thermal spectra from isolated
neutron stars, wavelength information, and thus energies
of excited states, are requisite. Jones et al. \cite{jones97} have shown
a way how to calculate excited states of small atoms in strong magnetic fields using the correlation function Monte Carlo method \cite{bernu88}.  The challenge remains to transfer their method to the DQMC simulations
presented in this paper, and to calculate excited states of large atoms in intense fields.

\section*{Acknowledgments}

This work was supported by Deutsche Forschungsgemeinschaft within the  
SFB 382 "Methods and Algorithms for Simulating Physical Processes on High-Performance Computers"
at the Universities of T{\"u}bingen and Stuttgart.

\end{document}